\documentclass[twocolumn,showpacs,preprintnumbers,amsmath,amssymb]{revtex4}

\usepackage{graphicx}
\usepackage{dcolumn}
\usepackage{bm}
\input{psfig}

\begin{document}


\title{Influence of the passive region on Zero Field Steps 
for window Josephson junctions}
\author{A. Benabdallah}
\email{abenab@mpipks-dresden.mpg.de}
\affiliation{Max-Planck-Institut f\"ur Physik komplexer Systeme,
 N\"othnitzer Stra\ss e 38, D-01187 Dresden, Germany}

\author{J.\ G. Caputo}
\email{caputo@insa-rouen.fr}
\affiliation{
Laboratoire de Math\'ematique, INSA de Rouen
B.P. 8, 76131 Mont-Saint-Aignan cedex, France}
\affiliation{
Laboratoire de Physique th\'eorique et modelisation
Universit\'e de Cergy-Pontoise and C.N.R.S. }
\date{\today}

   \begin{abstract}

We present a numerical and analytic study of the influence
of the passive region on fluxon dynamics in a window junction.
We examine
the effect of the extension of the passive region and its
electromagnetic characteristics, its surface inductance
$L_I$ and capacitance $C_I$.
When the velocity in the passive region $v_{I}$ is equal to the
Swihart velocity (1) a one dimensional model describes well
the operation of the device. When $v_{I}$ is different from
1, the fluxon adapts its velocity to $v_{I}$. In both cases
we give simple formulas for the position of the limiting
voltage of the zero field steps. Large values of
$L_I$ and $C_I$ lead to different types of solutions which
are analyzed.

\end{abstract}

\maketitle



\section{ Introduction}

The present design of low temperature superconducting devices
based on semi-conductor technology enables to integrate Josephson
junctions into superconducting strip-lines to realize complex
devices. A simple example is
the window design shown in Figure 1 where a single rectangular
junction is surrounded by a uniform passive region. In other
systems like integrated receivers \cite{as01}, the local flux flow 
oscillator is coupled directly to a strip-line containing a few junctions
which will realize the mixing of the local signal with the external signal
coming from an antenna.

The electrodynamic behavior of such window junctions is given by 
the phase difference between the top and bottom superconducting
layers. It obeys an inhomogeneous sine-Gordon equation which 
is equivalent to a homogeneous sine-Gordon equation
for the junction domain coupled to a wave equation for the
passive region.\\
In the static case, a small passive region causes a rescaling of the Josephson
length $\lambda_J$ into $\lambda_{\rm eff} > \lambda_J$ \cite{cfv96} and
this renormalization was seen in the experiments by the group of Ustinov 
\cite{fwu01}. On the other hand a large passive region causes the destruction
of the kink due to the finite length of the junction \cite{cfd94}.

In the dynamical case two limiting systems can be considered, 
first the passive region can be present only along the longitudinal direction
of the junction. Then one can assume a transverse profile that propagates in the
long direction. Using these ideas, Lee et al derived 
the dispersion relation for linear superconducting strip-lines \cite{Lee,lb92}.
The motion of kinks in such a window junction was studied by one of the authors
using periodic boundary conditions in $x$ \cite{flckc00}. It was shown that the
speed of the kink depends on the extension of the passive region and that 
Cerenkov resonances occur between the cavity modes and the kink. These 
show up as steps in the zero field steps (ZFS) in the IV characteristics. 
The experiments conducted in this geometry, for a fixed set of electric
parameters \cite{bmu96} confirm these findings.

In the second limiting case, which we will call the 1D model, 
the passive region exists only at each end of the junction as shown 
in the bottom left panel of Figure 1. Then the passive region acts
in a different way, and fixes the speed of the kink \cite{bc00}.
In \cite{bc00} we also studied the 2D problem i.e. a homogeneous 
passive region surrounding the junction, using an adapted numerical 
method, the finite volume approximation  together with soliton perturbation
theory. The 1D model showed that interfaces can act as fluxon 
traps. Numerical results in the two dimensional case indicated that the 
sheet inductance miss-match $L_I/L_J$ between the junction and the 
passive region is the main cause of instability of the kink motion.

Here we confirm these results, give the details of the characteristic
IV curves and ZFS for the 1D and 2D models and compare these two situations
for different geometrical and electrical parameters.
We also give a complete derivation
of the continuum model used in \cite{bc00} from the Resistive Shunted Junction 
(RSJ) approximation and explain discreteness effects observed in the solution of 
the 1D model.
For the 2D model we justify the fact that the kink adapts
its velocity to the one in the passive region.
Finally we provide insight into the kink stability for large sheet
inductance in the passive region $L_I$ or capacitance per unit surface
$C_I$. In the 2D case we give specifically the different IV curves
which can be seen for the different values of the sheet inductance in
the passive region $L_I$.

The paper is organized as follows. After deriving the continuum equations
for the RSJ model in section 1, we introduce the 1D model and study its
zero field steps in section 2. In Section 3 we study the ZFS for the
2D model and compare them to the ones for the 1D model.
We give our concluding remarks in section 4.
\begin{figure} [h]
\centerline{\psfig{figure=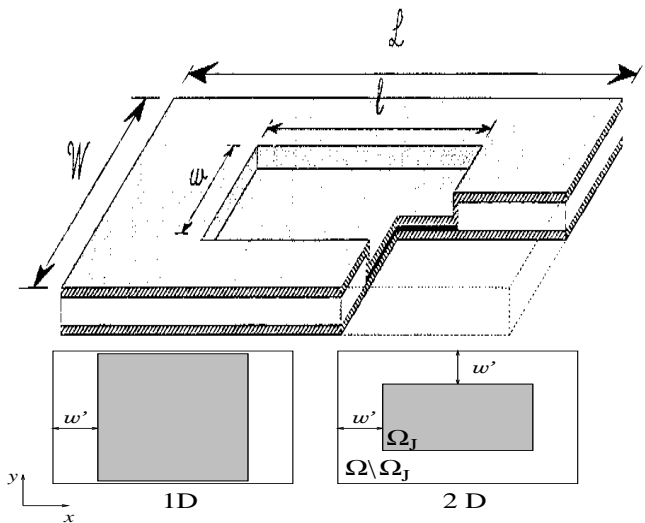,angle=-90,width=8.5cm,height=7cm}}
\caption{A view of a window Josephson junction. The
bottom left panel shows a schematic top view. For the system shown
on the right the linear region exists only on the left and right sides
of the junction.}
\label{geo}
\end{figure}

\begin{figure} [h]
\centerline{\psfig{figure=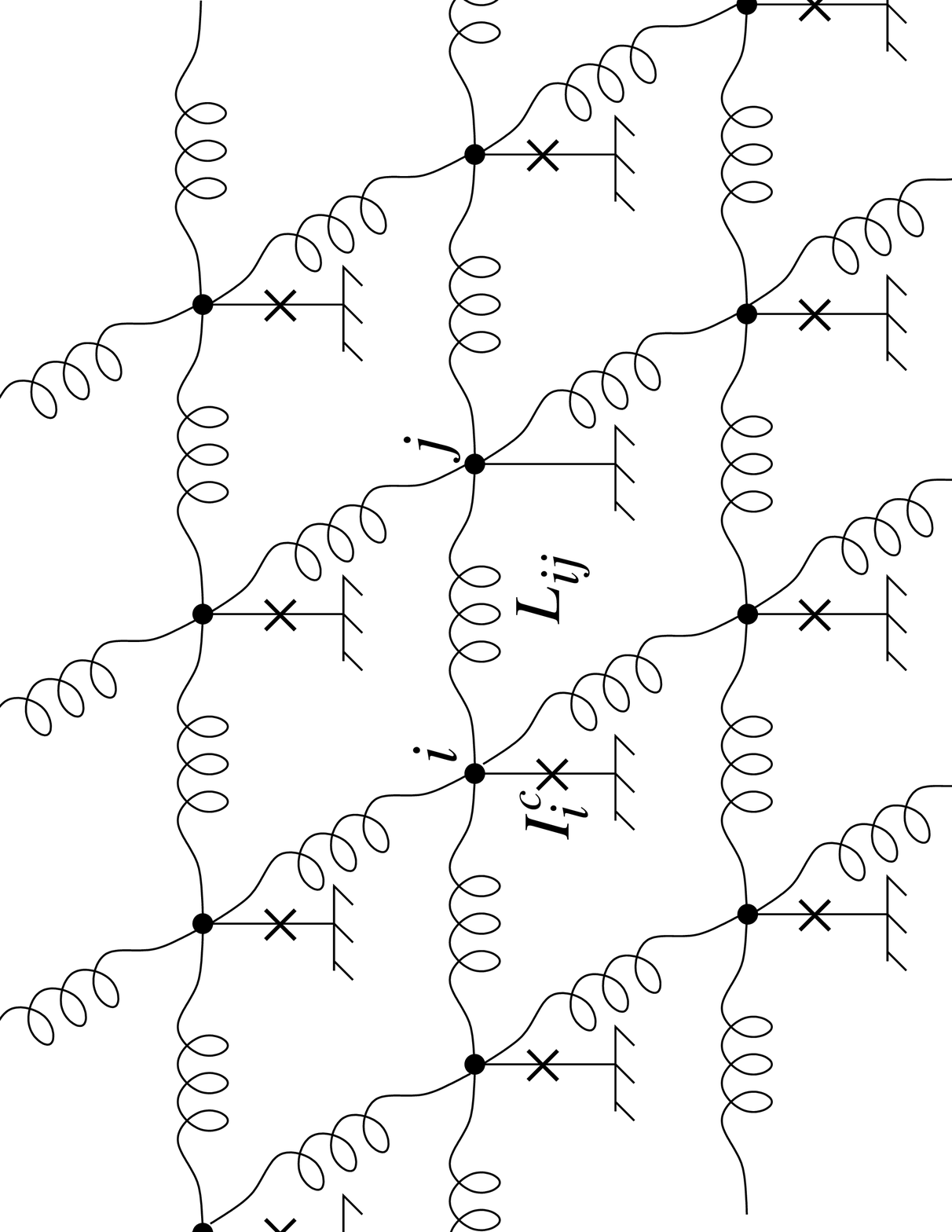,angle=-90,width=7cm,height=5cm}}
\caption{Portion of equivalent circuit model corresponding to the window
junction displayed in Fig. \ref{geo}. Crosses represent tunnel elements. The 
critical curent $I_i^c$ is nonzero only inside de window}
\label{rsj}
\end{figure}

\section{ The model}

\subsection{RSJ model for window Josephson junction}

A simple mathematical model of the window Josephson junction is to describe
each superconductor by an array of inductances $L$ (see Fig. \ref{rsj}). The coupling elements
between two adjacent nodes in each array are, a capacitor $C$, a resistance $R$
and a Josephson current $I_c$ \cite{Likharev, Devoret}. The Kirchoff laws at 
each couple of nodes $(i_b,i_t)$ in the bottom and top superconducting layers can be 
combined to give the relation expressing the conservation of currents 
at node $i$ in the Josephson junction
\begin{equation}
\label{eqjjf}
C_J \ddot \Phi_i + \sum_j  \frac {\Phi_i - \Phi_j}{L_{ij}} + I_i^c \sin \frac {\Phi_i}
{\Phi_0} +  \frac {\dot \Phi_i}{R_i} = 0 \; ,
\end{equation}
and in the passive region
\begin{equation}
\label{eqi}
C_I \ddot \Psi_i + \sum_j \frac {\Psi_i - \Psi_j}{L_{ij}} = 0 \; ,
\end{equation}
where $\Phi = \Phi_t - \Phi_b$ (resp. $\Psi = \Psi_t - \Psi_b$), 
is the phase difference between the two superconductors 
in the junction (resp. passive) part,  the summation $\sum_j$ is applied to 
the nearest neighbors. 
Also note that equations (\ref{eqjjf}) and (\ref{eqi}) are discretisations of Maxwell's equations 
(wave equation part) and Josephson constitutive equations (sinus term), assuming an electric
field normal to the plates, a magnetic field in the junction plane and perfect symmetry
between the top and bottom superconducting layers. We can now obtain the model
in the continuum limit, more suitable for analysis.

\subsection{Continuum limit}

The continuum version of the system (\ref{eqjjf})-(\ref{eqi}) 
can be derived by introducing the following quantities
per  unit area ($a^2$) of elementary cells 
of length $a$.
\begin{equation}
\label{q1}
\overline {C_{I,J}}=\frac {C_{I,J}} {a^2},~~~j_c= \frac {I_i^c} {a^2},~~~\overline{r} = R a^2.~~~
\end{equation}
We normalize the phases by the flux quantum $\Phi_0$,
\begin{equation}
\label{q4}
\phi_i=\frac {\Phi_i}{\Phi_0} \quad,\quad
\psi_i   = \frac {\Psi_i}{\Phi_0} \; ,
\end{equation}
and introduce the Josephson characteristic length $\lambda_J= a \tilde {\lambda_J}$
\begin{equation}
\label{q5}
\lambda_J^2=\frac {\phi_0}{2\pi j_cL_J} = \frac {\Phi_0}{j_cL_J}=
\tilde {\lambda_J}^2 a^2 \; .
\end{equation}
Notice that in this 2D problem the inductance associated to each cell
is equal to the branch inductance $L_J$ (resp. $L_I$) in the
junction (resp. passive region). This is not the case for the 1D model
for which these inductances are in series, giving a total
inductance proportional to the mesh size $dx$.

We  substitute relations (\ref{q1}) - (\ref{q5}) in (\ref{eqjjf}) and 
(\ref{eqi}) and obtain in the  junction
\begin{equation}
\label{eqjjf1}
a^2\Phi_0\overline {C_J} \ddot \phi_i +  \Phi_0 \sum_j \frac {\phi_i - \phi_j}{L_{ij}} 
+ a^2 j_c  \sin \phi_i + \frac {a^2}{\overline{r}} \Phi_0 \dot \phi_i = 0 \; ,
\end{equation}
{\rm and in the passive part}
\begin{equation}
\label{eqi1}
a^2\Phi_0 \overline {C_I} \ddot \psi_i + \Phi_0\sum_j \frac{\psi_i - \psi_j}{L_{ij}}=0 \; .
\end{equation}
After simplification of Eqs. (\ref{eqjjf1})-(\ref{eqi1}) and using Eq. (\ref{q5}) 
we obtain 
\begin{equation}
\label{eqjjf2}
\overline {C_J}L_J \ddot \phi_i +  \sum_j \frac {\phi_i - \phi_j}{a^2}
+ \frac {1}{\lambda_J^2}\sin \phi_i + \frac {L_J}{\overline{r}}\dot \phi_i = 0 \; ,
\end{equation}
\begin{equation}
\label{eqi2}
\overline {C_I} L_I \ddot \psi_i + \sum_j \frac{\psi_i - \psi_j}{a^2}=0 \; .
\end{equation}
\noindent Now we take the limit $a \rightarrow 0$, so that\\
$\displaystyle{\sum_j \frac {\phi_i - \phi_j}{a^2} \rightarrow -\Delta \phi}$ \quad and \quad 
$\displaystyle{\sum_j \frac{\psi_i - \psi_j}{a^2} \rightarrow -\Delta \psi}$.\\ 
and obtain the following system of partial differential equations 
\begin{equation}
\label{eqc1}
\overline{C_J} L_J \frac {\partial^2 \phi}{\partial t^2} - \Delta\,\phi
+ \frac 1 {\lambda_J^2}\sin \phi 
+ \frac {L_J}{\overline{r}} \frac {\partial \phi}{\partial t}= 0 \; ,
\end{equation}
\begin{equation}
\label{eqc2}
\overline{C_I} L_I \frac {\partial^2 \psi}{\partial t^2} - \Delta\,\psi=0 \; .
\end{equation}
We introduce the plasma frequency in the junction $ \omega_J^{-2} = 
\lambda_J^2 v_{I}^{-2}\; ,$
where $v_{I}^{-2} = \overline{C_J} L_J$ is the Swihart velocity  \cite{Swihart}.
Similarly we define the electromagnetic wave frequency in the passive part
$\omega_I^{-2} = \lambda_J^2 v^{-2} \;, $ where  $v^{-2} = \overline{C_I} L_I$.
In the following we fix the couple $(L_J, C_J)$ and $\omega_J$ and 
vary the inductance $L_I$ and capacity $C_I$ in the passive region. \\
We normalize equations (\ref{eqc1}) - (\ref{eqc2}) in space by 
 $\lambda_J$ and in time  by $\omega_J^{-1}$ and obtain
in the junction  ($\Omega_J$)
\begin{equation}
\label{eqc3}
\frac {\partial^2 \phi}{\partial t^2} - \Delta\,\phi
+ \sin \phi + \alpha  
\frac {\partial \phi}{\partial t}= 0 \; ,
\end{equation}
{\rm and in the passive region} $(\Omega \backslash \Omega_J)$
\begin{equation}
\label{eqc4}
\left(\frac {\omega_J} {\omega_I}\right)^2 \frac {\partial^2 \psi}{\partial t^2} 
- \Delta\,\psi=0 \; ,
\end{equation}
where the dimensionless parameter $\alpha= \frac {\tilde {\lambda_J}}{R}
\sqrt{\frac {L_J}{\overline{C_J}}} $
is a damping coefficient which depends on the spatial discretization 
via $\tilde {\lambda_J}$.\\

The boundary conditions of equations (\ref{eqc3}) and (\ref{eqc4}) are of 
inhomogeneous Neumann type which physically indicates a lateral injection 
of current or/and an external magnetic field  
\begin{equation}
\label{bc}
\frac 1 {L_I} \nabla \psi \cdot  n = J_{ext}\;\; {\rm on}\;\; \partial \Omega \; ,
\end{equation}
where $n$ is the exterior normal. 
\noindent To this we add the interface conditions
   for the phase and its normal gradient,
   the surface current on the junction boundary $\partial \Omega_J$
   \begin{equation}  \label{ic}
   \psi = \phi ~~~ {\rm and} ~~~
   {1 \over L_I} {\partial \psi \over \partial n} = {\partial \phi \over
   \partial n}
   \end{equation}
   
   \noindent Note that the jump condition (2nd relation in (\ref{ic}))
   can be obtained by integrating (\ref{eqc3}) - (\ref{eqc4}) on a small surface
   overlapping the junction domain $\Omega_J$. \\
   In the rest of the paper we assume a rectangular window
   of length $l=10$ and width $w=1$ embedded in a
   rectangular passive region of extension $w'$ as shown in
   Figure 1. We will not consider the
   influence of an external magnetic field and will assume
   the external current feed to be of
   overlap type so that the boundary conditions (\ref{bc}) become
   \begin{eqnarray}
  \nonumber
  \frac 1 {L_I}\frac {\partial \psi}{\partial y}(x,0) & = &
  -\frac I {2{\cal L}} , ~~~
  \frac 1 {L_I}\frac {\partial \psi} {\partial y}
  (x,{\cal W})= \frac I {2{\cal L}}  , \\ 
  \frac {\partial \psi}{\partial x}(0,y)
  & = & \frac {\partial \psi} {\partial x}({\cal L},y)= 0 ,  \label{rbc}
  \end{eqnarray}
  where ${\cal L} = 2w' + l \quad {\rm and}
  \quad {\cal W}=2w' + w$.\\

   We assumed throughout the study a small damping $\alpha=0.01$ which is
   typical of under damped Josephson junctions.

   \section{A 1D window Josephson junction}

\subsection{The model}

We first consider the simplified situation where the passive region
is present only at the two lateral ends of the device as shown in the bottom
left panel of Fig. 1. For this system it is possible to derive the
effective 1D equation \cite{bc00}
\begin{eqnarray}
\nonumber
C(x)\varphi_{tt}-\partial_x\left(\frac 1{L(x)}\varphi_x\right)
 +  \epsilon(x)(&\sin \varphi  +  \alpha \varphi_t)  -
\gamma(x)  = 0 \\  
 &{\rm for }~~ x \in ]0,{\cal L}[ , \label{eq1d}
\end{eqnarray}
where $\varphi$ is the phase averaged in y across the
junction and $\gamma(x) = \frac{I}{L(x) s}$ where $s$ is the junction
area.\\
In equation (\ref{eq1d}) $C,L$ and $\epsilon$ are
discontinuous functions of $x$ defined by
\begin{equation}
(C(x),L(x),\epsilon(x))=\left\{
\begin{array}{cc}
(1,1,1)& \;\;{\rm if }\;\;x \in \rbrack w',w'+l\lbrack  \\
\\
(C_I,L_I,0)&\;\;{\rm if }\;\;x \in \rbrack 0,w'
\lbrack \cup \rbrack w'+l,{\cal L} \lbrack  \; ,
\end{array}
\right.
\end{equation}
The boundary conditions (\ref{rbc}) reduce then to
\begin{equation}
\label{bc1d}
\varphi_x(0)= \varphi_x({\cal L})=0 \; .
\end{equation}

To compare different values of $w'$ we normalize the 
current $I$ by the maximum current that the
junction can carry $I_{\rm max}=s$. To simplify the calculations
we have assumed a constant current density $\gamma$.

We choose the values of the inductance $L_I$ and capacity
$C_I$ according to two strategies. First we assume
$L_I \, C_I=v_{I}^{-2}$ is constant
which gives a uniform velocity in the passive part. In a second set
of calculations we
vary $v_{I}$ by moving orthogonally to the set of hyperbolas
in the $(L_I,C_I)$ plane.

The numerical method to obtain the (IV) curve is to start with 
a static kink like solution obtained by solving the static problem
derived from (\ref{eq1d}) and to increase progressively the current to
obtain a moving fluxon. After obtaining a stable solution
in time, we compute the voltage in the center of the junction
by
\begin{eqnarray}
\nonumber
V\equiv\frac 1{T_2-T_1}\int_{T_1}^{T_2}\varphi_t(w'+l/2,t) 
\;{\rm d}t = \\  \frac {\varphi(w'+l/2,T_2) - 
\varphi(w'+l/2,T_1)}{T_2-T_1} \, , \label{eqv} \end{eqnarray}
where the times $T_1$ et $T_2$ are given such that the voltage 
is stable.

\subsection{Discreteness effects}

\begin{figure} [h]
\centerline{\psfig{figure=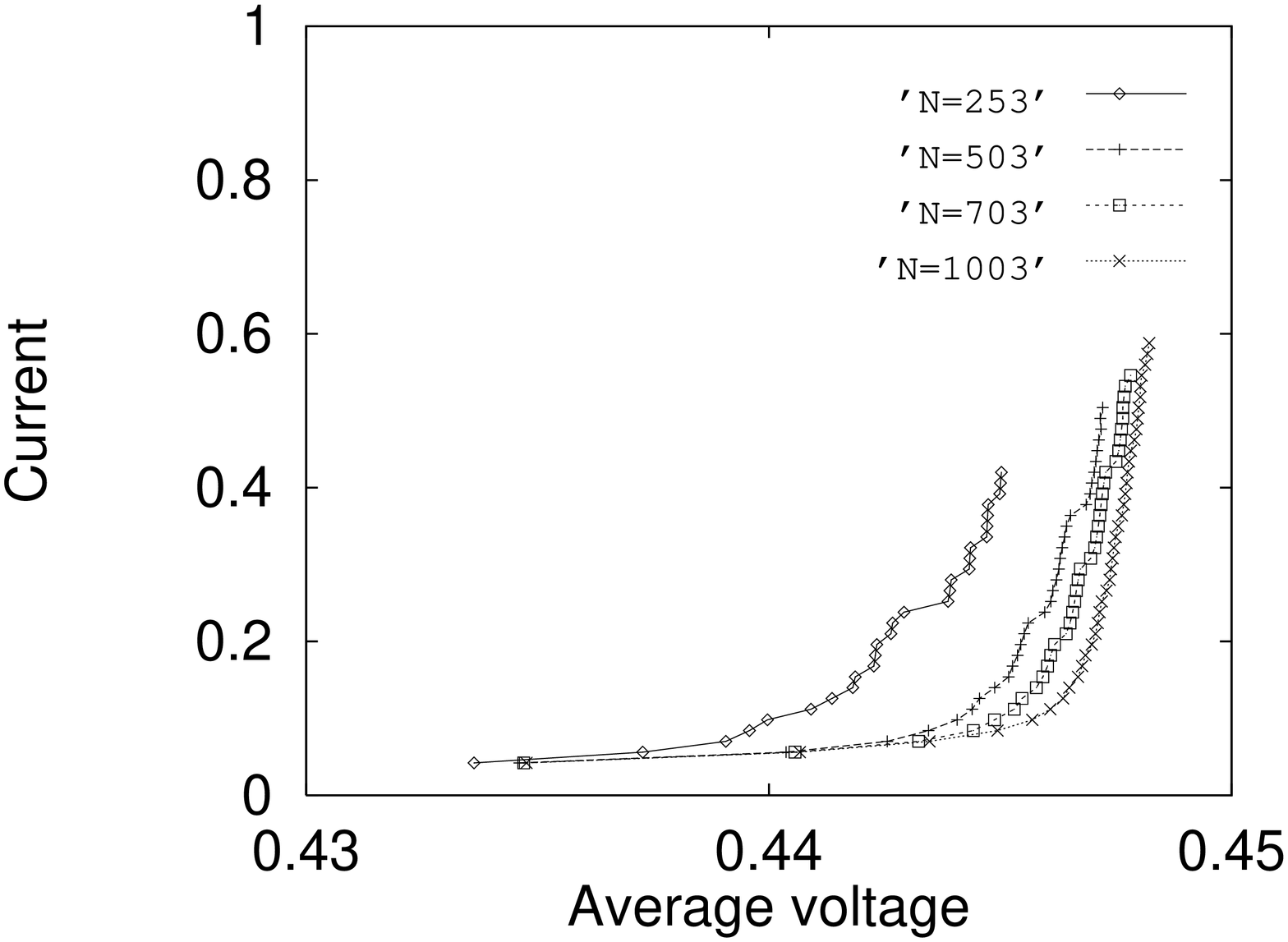,angle=-90,width=8cm,height=6cm}}
\centerline{\psfig{figure=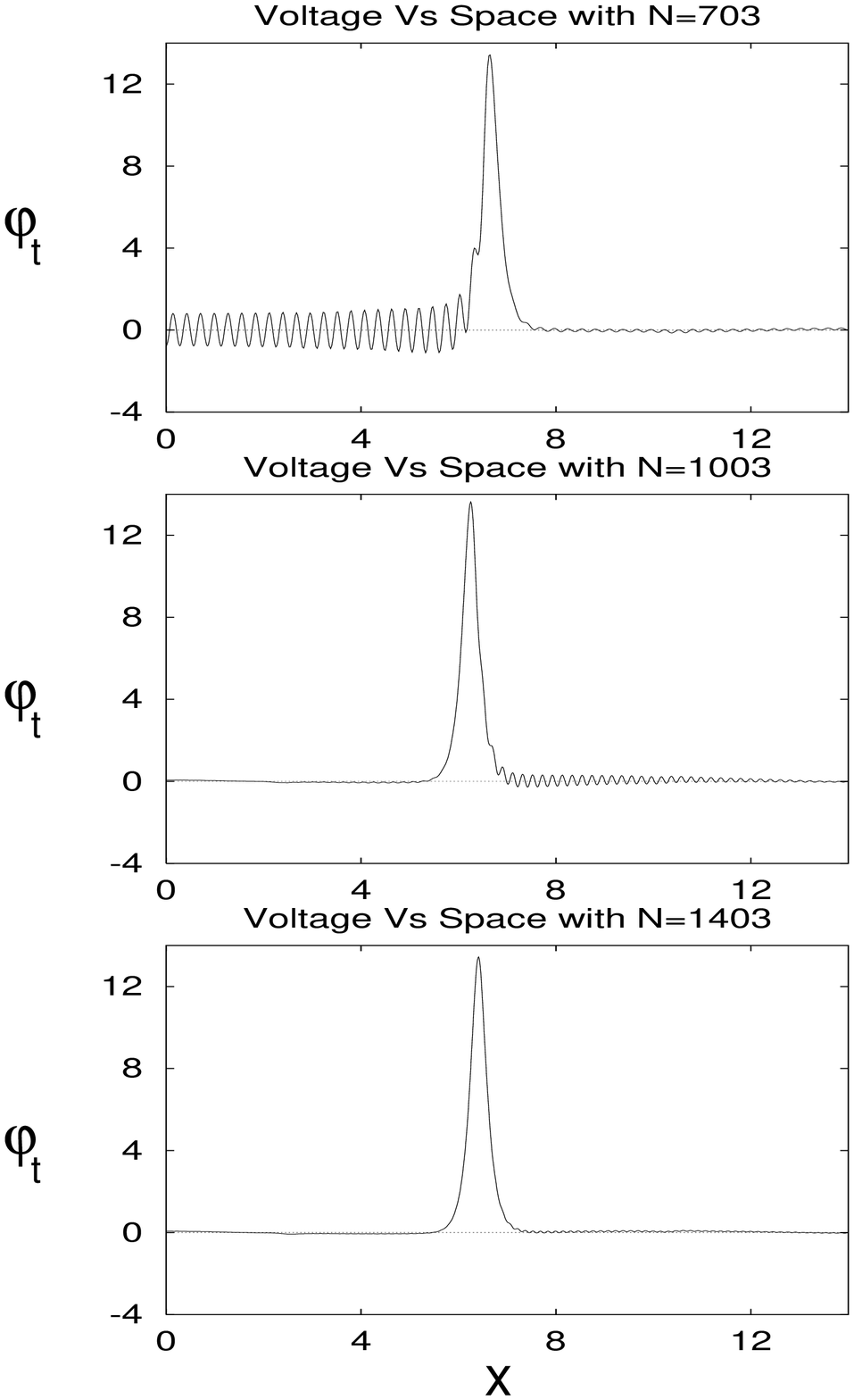,angle=0,width=9cm,height=6cm}}
\caption{1D dynamics. The top panel shows the current voltage characteristic 
for $w'=2$ and various number of discretization points. The bottom panel
shows the corresponding instantaneous voltage $\varphi_t(x) $ for a given time
for three different values of the discretization $N = 703, 1003$ and $1403$. 
The current density is equal to $0.08$.}
\label{vxt}
\end{figure}

\noindent In this section we consider a window junction with 
homogeneous electric properties ($L_I=C_I=1)$ and fix the width of 
the passive part $w'=2$.
We have chosen the number of discretization points in the $x$
direction as $N = 253, 503, 703$ and $1003$.
In this case we compute the IV characteristic as shown
in the top panel of Fig. \ref{vxt}. 

For small values of $N$, we observe resonances in the IV curve
which decrease and disappear as the number of mesh points is
increased from $N=253$ to 703. These disappear altogether for
$N=1003$. To see if these resonances are due to the presence of the
passive region, we have calculated a IV curve for a homogeneous
junction ($w'=0$) and indeed they are absent. To understand the
mechanism of this fine structure in the IV curve, we have plotted on
the bottom panel of Fig.\ref{vxt} the instantaneous voltage $\phi_t$ as a function of $x$
for a fixed time and $N=703, 1003$ and $1403$. From the comparison of the
two panels of Fig. \ref{vxt}, it is clear that the radiation is
responsible for the fine structure. In fact the wave length of the excited
radiation corresponds to a standing mode $V= 2~ n \pi / 4$ with $n\approx 10$.

\begin{figure} [h]
\centerline{\psfig{figure=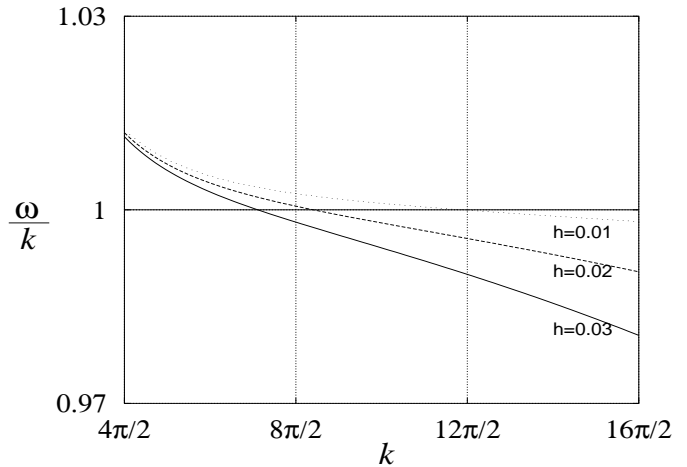,angle=-90,width=10cm,height=7cm}}
\caption{Phase velocity of linear waves for the discrete model 
$v_{\varphi}$ vs. $k$ for three values of the step 
discretization $h=0.01,~0.02$ and $h=0.03$}
\label{vph}
\end{figure}

Let us now explain how such radiation is excited by the kink as it crosses an
interface from a linear to a nonlinear medium. Fig. \ref{vph} shows the
phase velocity $v_\phi = \omega / k$ for the continuum sine-Gordon system
(where $\sin\phi\approx \phi$) and the discrete system used numerically for which
the second derivative is approximated by a three points difference
$\phi_{xx} =(\phi_{n+1} + \phi_{n-1} -2 \phi_{n})$. For the former
$v_\phi =\sqrt{1+1/k^2}~~ >~~1$ so that nonlinear modes cannot be excited
via the long wave short wave Benney resonance mechanism \cite{degm82}. This is
not true for the discrete system for which
$$v_\phi ={1 \over k}\sqrt{1+{4\over h^2} \sin^2(kh/2)} $$ 
\noindent can be smaller than 1 for $h\neq 0$ and small wave lengths.
As expected the threshold $k$ decreases as is increased. Notice that the
scale in $k$ corresponds to $9 \times  \pi / 2  \leq k \leq 12 \times \pi / 2$.

We then explain quantitatively the resonances in the IV curve shown in
Fig. \ref{vxt} by the fact that as the kink velocity increases, it can
lock with a given cavity mode. For example $N=703$ gives voltage (resp. velocity) 
steps $V = 0.4457, 0.44645, 0.447, 0.4475$ (resp. $v = 0.9935, 0.9952, 0.996, 0.997)$
which correspond to the cavity mode index $n = 14, 13, 12$ and $11.$ For $N = 503$ we observe
steps at voltages $V=0.4451, 0.446, 0.4467$ (resp. $v = 0.9922, 0.99426, 0.9958)$
for which $n = 11, 10$ and $9$ respectively. The lower values of $n$ explain the larger 
amplitudes observed in the radiation.

An intrinsic fine structure in the IV characteristics has been seen in
experiments with window junctions with a lateral passive region. For example
Thyssen et al. in \cite{Capri} show a slight shift in the position (and therefore
limiting fluxon velocity) of the IV curves for different values of $w'$. Ustinov et
al \cite{bmu96} also display IV curves which depend on $w'$ and large substructures.
Recent numerical work \cite{flckc00} on a Josephson window junction with a 
lateral passive region and periodic boundary conditions along the propagation 
direction confirm these resonances due to Cerenkov radiation between a 
soliton travelling faster than
$v=1$ and the radiation of phase speed $v_\phi =\sqrt{1+1/k^2} > 1$.
Here we show that when the passive region exists only in the propagation
direction, resonances disappear at large resolutions. This can be expected
from the calculation of the emitted power of radiation for a kink as it 
crosses an interface \cite{kc91,km89}. This quantity drops fast to zero as the
speed approaches 1.


In the rest of the study we have chosen
$N=1403$ so that numerical resonances are eliminated from the
IV characteristic.

\subsection{Zero field steps (ZFS)}
  
We now proceed to give an estimate of the time average 
of the voltage for the window junction. For this,
we compute the velocity in the junction using the
McLaughlin-Scott soliton perturbation equations \cite{ms78} and get
for the kink velocity
\[
v = \frac 1 {\sqrt {1 + \left(\frac {4 \alpha}{\pi \gamma}\right)^2}} \; .
\]
When the current $\gamma$ increases and becomes larger than
$\alpha$, the velocity $v$ gets close to 1.\\
In the passive part the velocity $v_{I}$ is given by
(\ref{eq1d}) 
\[
v_{I}= \frac 1 {\sqrt {L_I\,C_I}} \; ,
\]
so that the average of the voltage in time in the limit of large
current can be approximated by 
\begin{equation}
\label{v1d}
V_{\rm 1D} \equiv <\varphi_t>=
{{\Delta \varphi} \over {\Delta t}}
= \frac {2\pi}{2w'/v_{I} + l/v} \;  ,
\end{equation}
\noindent where the denominator is just the time  take
by the kink (phase jump of $2 \pi$) to travel across the
device. We now confirm this estimation first
when the electric properties of the device are homogeneous
and then when they are different in the junction and the passive
region so that $v_I \neq 1$.

\subsection{Influence of geometry}

\begin{figure} [h]
\centerline{\psfig{figure=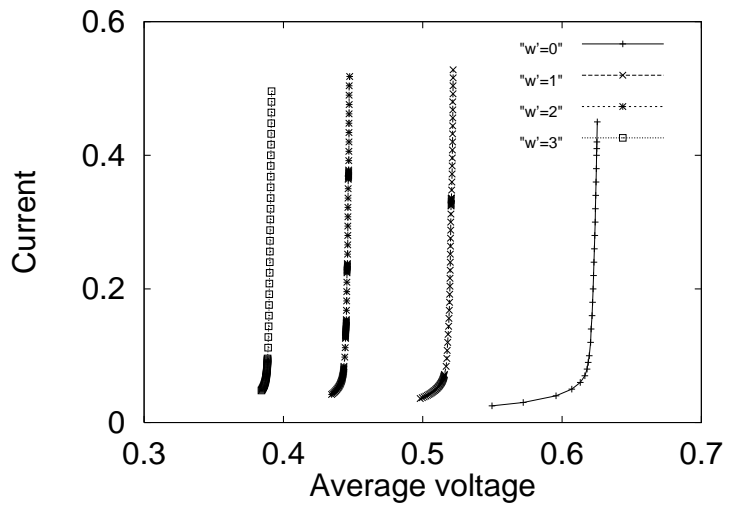,angle=-90,width=10cm,height=7cm}}
\caption{The 1D current voltage characteristic for $L_I=C_I=1$ and various 
extensions of the passive region $w'$.}
\label{iv3}
\end{figure}

First we consider a situation where the electric properties 
are homogeneous in the whole device so that $(L_I=C_I=1)$ 
and change the extension of the passive region $w'$.\\
In Fig. \ref{iv3} we plot the (IV) characteristic 
curves for four values of $w'=0, 1 , 2$ and $3$. This figure.
show that the position of the ZFS moves towards the left when $w'$ 
increases. This agrees with formula (\ref{v1d}). Table (\ref{tab1}) 
reports the limiting values of the zero field 
steps and the estimates from (\ref{v1d}), which are in excellent agreement.
For large value of $w'$ (typically $w'=12$), the kink becomes unstable.
This seems to be due to the fixed points that exist on each
junction/passive region interface \cite{bc00}. The kink can be trapped
easier by one of them as $w'$ increases because the driving force due to 
the current gets weaker.

\begin{table}[htbp]
\caption{positions of the ZFS for the 1D window
model homogeneous $(L_I=C_I=1)$ together with the estimates (\ref{v1d}).}
\vspace{0.2cm}
\begin{center}
\footnotesize
\begin{tabular}{||c|c|c||}
\hline  $w'$ & $V_{\rm 1D}$ & $V_{\rm num}$1D \\  
\hline  $0$ & $0.628$ & $0.625$ \\
\hline  $1$ & $0.523$ & $0.521$ \\
\hline  $2$ & $0.448$ & $0.447$ \\
\hline  $3$ & $0.392$ & $0.391$ \\ \hline
\end{tabular}
\vspace{0.2cm}
\end{center}
\label{tab1}
\end{table}

\subsection{Influence of the electrical parameters}

\begin{figure} [h]
\centerline{\psfig{figure=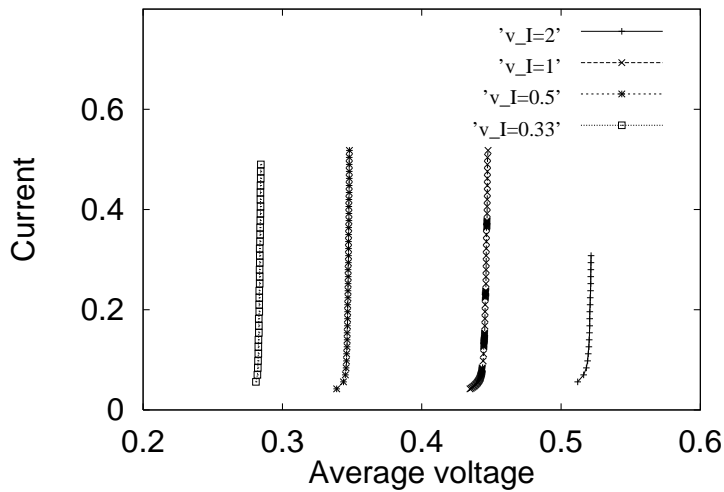,angle=-90,width=10cm,height=7cm}}
\caption{The 1D Current voltage characteristic for $w'=2$ and various
velocity in the passive region. The velocities $v_I$ are indicated
in the top right corner of the picture.}
\label{iv4}
\end{figure}

The second case tested is when the velocity in the passive region
$v_{I}\neq 1$. To simplify things we fix $w'$ to $2$ in all 
the presented runs.
We show in Fig. \ref{iv4} the IV characteristics for 
four values of the velocity in the passive 
region $v_{I}= 0.33, 0.5, 1$ and $2$.  
Notice that the positions of the ZFS of the 1D effective model follow
(\ref{v1d}) as shown in table (\ref{tab2}).
Fig. \ref{iv4} also shows that the position of the ZFS moves from the
right to the left as $v_{I}$ decreases from $2$ to $0.33$.
When $v_{I}$ goes to zero so that $L_I$ or $C_I$ goes to infinity, we
expect the ZFS to disappear. We observe regions of instability for 
$L_I >> 1$ or $C_I >> 1$ and 
the zero field steps exist only in small intervals in the 
$(L_I ,C_I) $ plane. In Fig. \ref{zs} we present a detailed
numerical exploration of the $(L_I,C_I)$ plane. The hyperbolas corresponding
to $L_I \, C_I={v_{I}}^{-2}$ constant are shown. We indicate by the 
symbols $(*),$ $(\times)$ and $(+)$ the points where the ZFS exist. 
We note that for these points
the solution of the window 1D problem is a kink. The region of
stability is concentrated along the diagonal. Note that the results 
of Fig. \ref{zs} are based only on the numerics.\\ 
When $C_I$ or $L_I$ become large we obtain particular types of solutions which we
now discuss.

\begin{table}[!h]
\caption{positions of the ZFS for the 1D window
model together with their corresponding estimates (\ref{v1d}) for $w'=2$.}
\vspace{0.2cm}
\begin{center}
\footnotesize
\begin{tabular}{||c|c|c|c||}
\hline  $L_I\,C_I$& $v_{I}$&  $V_{\rm 1D}$ & $V_{\rm num} 1D$ \\
\hline  $0.25$    & 2 &  $0.523$ & $0.521$ \\
\hline  $1 $      & 1 &  $0.449$ & $0.448$ \\
\hline  $4 $      & 0.5 & $0.349$ & $0.348$ \\
\hline  $9 $      & 0.33 & $0.285$ & $0.284$ \\ \hline
\end{tabular}
\end{center}
\label{tab2}
\end{table}

\subsection {Solutions for large $L_I$ or $C_I$}

Instabilities of the kink motion occur when the extension of the
passive region $w'$ is large, typically $w' \approx 12$. In that 
case the current density becomes small and cannot drive the kink out
of the potential well created by the interface \cite{bc00}.
\noindent
Other instability factors are the electrical parameters $(L_I, C_I)$.
We have shown in \cite{bc00, theabdel} that the electrical parameters 
act differently from the geometry. For $C_I >> 1$ the solution 
is static while it is dynamic for $L_I >> 1$. In this latter case, kink 
motion is possible only in the junction.
 
\noindent
We now consider the disappearance of the Zero field steps in 
the $IV$ curves. This occurs for intermediate values of
$C_I$ and $L_I$ and gives rise to a large voltage.

\noindent 
Let us recall the equations  of the problem
\begin{equation}
\label{wj1d1}
\frac {\partial^2 \phi}{\partial t^2}
- \frac {\partial^2 \phi}{\partial x^2}+ \sin \phi = \gamma - \alpha
\frac {\partial \phi}{\partial t} \quad w' \le x \le w'+l \; ,
\end{equation}
\begin{equation}
\label{wj1d2}
C_I \frac {\partial^2 \psi}{\partial t^2}
- {1 \over L_I} \frac {\partial^2 \psi}{\partial x^2} = \gamma
~~~ 0 \le x \le w' ~~{\rm and}~~  w'+l \le x \le {\cal L} \; ,
\end{equation}
\noindent together with the interface condition at $x=w',w'+l$
$$\phi = \psi ~~~, {1 \over L_I} \frac {\partial \psi}{\partial x}
= \frac {\partial \phi}{\partial x}~~,$$
\noindent and homogeneous boundary conditions
$\frac {\partial \psi}{\partial x}|_{x = 0} =
\frac {\partial \psi}{\partial x}|_{x = {\cal L}}= 0$.

\noindent
In the limit of large voltages $V$
$$\frac {\partial \phi}{\partial t} \approx V ~~~ {\rm and}~~~
\frac {\partial \psi}{\partial t}\approx V$$
so that one can average the above equations which then reduce to
\begin{equation} \label{awj1d1}
- \frac {\partial^2 \phi}{\partial x^2} = \gamma - \alpha V
\quad w' \le x \le w'+l \; ,
\end{equation}
\begin{equation}
\label{awj1d2}
- {1 \over L_I} \frac {\partial^2 \psi}{\partial x^2} = \gamma
~~~ 0 \le x \le w' ~~{\rm and}~~  w'+l \le x \le {\cal L} \; ,
\end{equation}
\noindent where the voltage $V = \gamma {\cal L} /(\alpha l)$ can
be derived from the work equation (see Appendix).

The time dependent part of $\phi,\psi$ is now just $V t$ and the $x$
dependent part verifies a boundary value problem with 
$\psi_x|_{x=0,{\cal L}}=0$ and the above given interface
conditions. The solution is symmetric with respect to $x={\cal L}/2$
and we obtain for the left and right passive regions and the junction
\begin{equation}
\label{ps1l}
\psi_l = -\gamma L_I {x^2 \over 2} + V t,
\end{equation}
\begin{equation}
\label{ps1r}
\psi_r = -\gamma L_I {({\cal L}-x)^2 \over 2} + V t 
\end{equation}
\begin{equation}
\label{ph1}
\phi = -\gamma {w' \over l} x ({\cal L}-x) + \gamma {{w'}^2 \over 2}
(2 - L_I + {2w' \over l}) + V t
\end{equation}

\begin{figure}[h]
\centerline{\psfig{figure=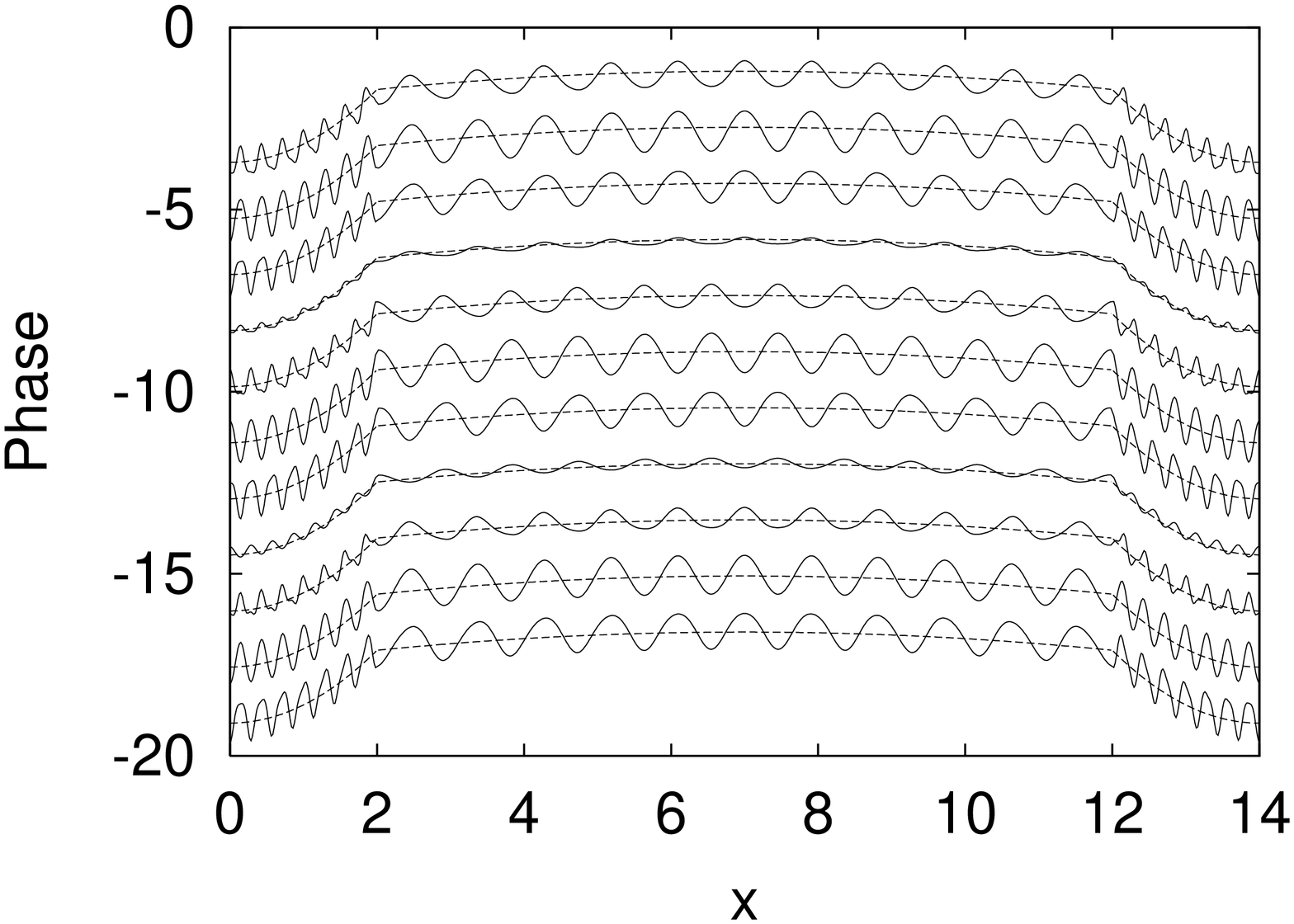,angle=-90,width=8cm,height=7cm}}
\centerline{\psfig{figure=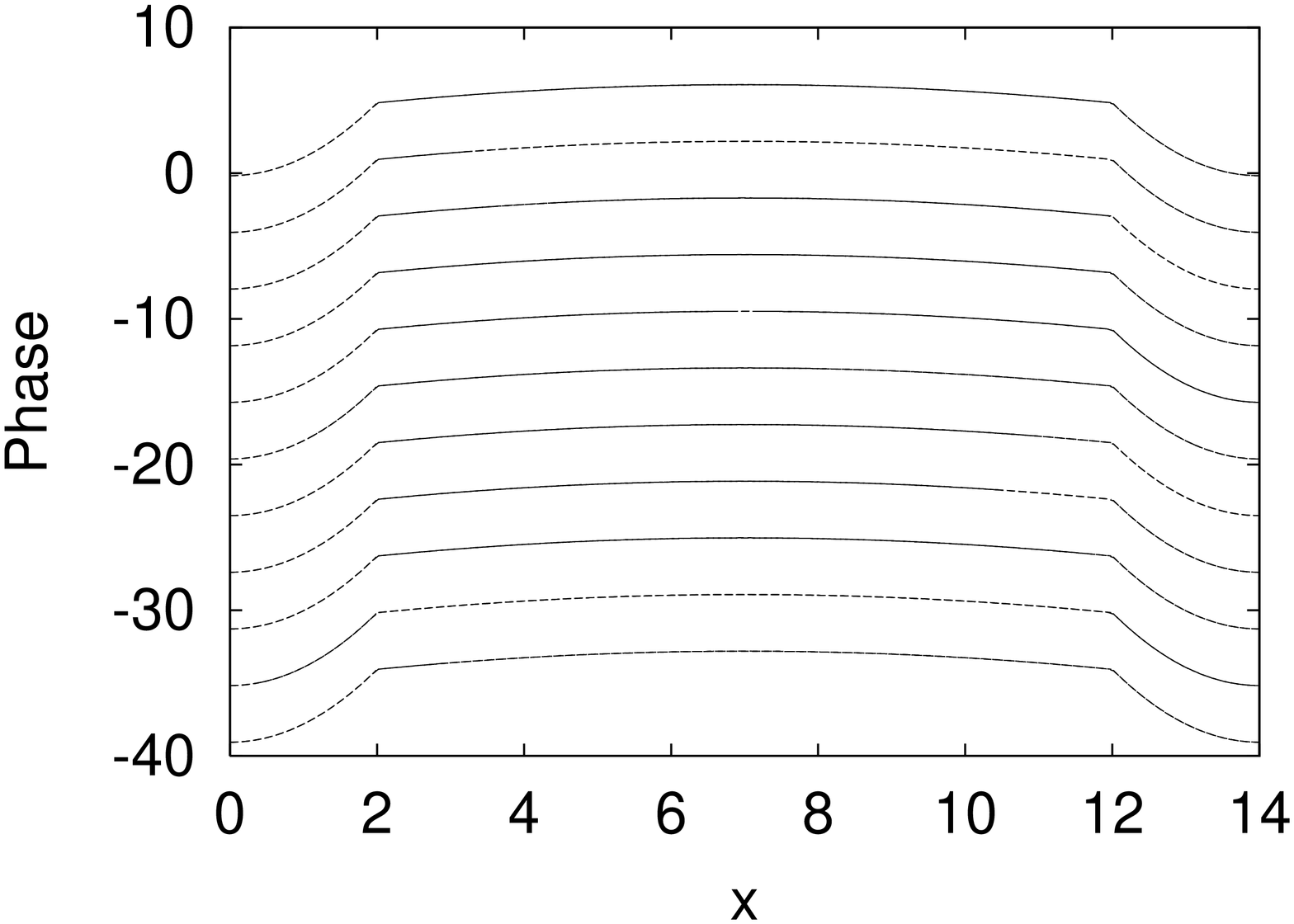,angle=-90,width=8cm,height=7cm}}
\caption{Phase $\varphi$ versus $x$ for eleven successive times separated
by $\Delta t=0.11$. The 
 parameters are $C_I=1$, $L_I=10$, $\gamma =0.1$ $V=14.$ (top panel) 
 and $\gamma =0.25$, $V=35.$ (bottom panel). 
 The numerical results are given in solid lines while the
 analytic expression (\ref{ps1l},\ref{ps1r},\ref{ph1}) are given
 in dashed lines.}
\label{largeli}
\end{figure}

The above given expressions give a very good approximation of the
phase computed numerically as can be seen in
Fig. \ref{largeli} which shows $\psi(x)$ and $\phi(x)$ for 
11 successive values of time. The electrical parameters are
$C_I = 1~~,L_I= 10$ and
the top panel corresponds to a small current density $\gamma = 0.1$
while the bottom panel is for $\gamma = 0.25$. The agreement for
the latter is excellent, showing complete overlap between the approximation
(\ref{ps1l},\ref{ps1r},\ref{ph1}) and the numerical solution. On the top panel
the approximations give the overall value but there are some oscillations
due to a resonance between the main frequency $V$ and a subharmonic.

It is possible to estimate the validity of the approximate solution
(\ref{ps1l},\ref{ps1r},\ref{ph1}) by considering perturbations around
it. This leads to the linearized equations for the perturbations
$\xi,\zeta$.
\begin{equation}
\label{eqlin1}
\frac {\partial^2 \xi}{\partial t^2}
- \frac {\partial^2 \xi}{\partial x^2}+ \xi \cos\phi +\alpha
\frac {\partial \xi}{\partial t}=0 \quad w' \le x \le w'+l \; ,
\end{equation}
 
\begin{equation}
\label{eqlin2}
C_I \frac {\partial^2 \zeta}{\partial t^2}
- {1 \over L_I} \frac {\partial^2 \zeta}{\partial x^2} = 0
~~~ 0 \le x \le w' ~~{\rm and}~~  w'+l \le x \le {\cal L} \;
\end{equation}
\noindent with the interface condition at $x=w',w'+l$
$$\xi = \zeta ~~~, {1 \over L_I} \frac {\partial \zeta}{\partial x}
= \frac {\partial \xi}{\partial x}~~,$$
\noindent and boundary conditions $\frac {\partial \zeta}{\partial x}|_{x = 0} =
\frac {\partial \zeta}{\partial x}|_{x = {\cal L}}= 0$.

We then assume a uniform time dependence $e^{-\alpha t/2 + i \omega t}$ and 
obtain a spectral problem which can be solved. We can neglect
the damping because it is small. 

In the particular case presented in the top panel of Fig. \ref{largeli}
the voltage is $V=14$ and the subharmonic observed is $V/4= 3.5$. This
corresponds to the mode $n_J=11$ in the junction and $n_I=7$ in each lateral
passive region of extension 2, taking into account the inductance $L_I=10$.

\begin{figure}[h]
\centerline{\psfig{figure=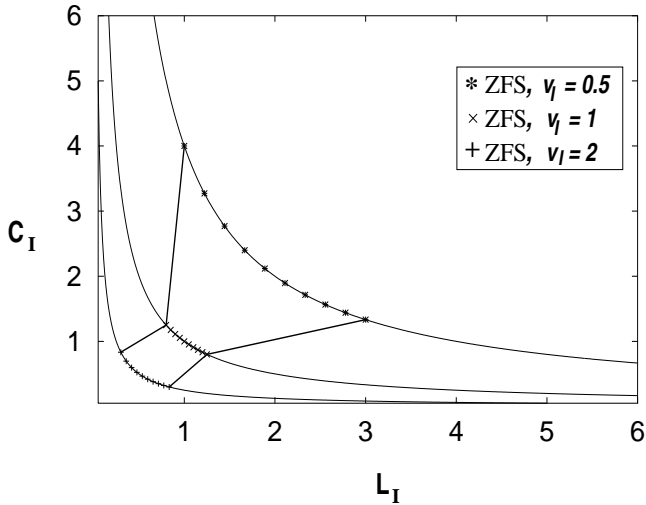,angle=-90,width=9cm,height=7cm}}
\caption{Parameter plane $(L_I, C_I)$ showing the regions
   of existence of zero field steps (ZFS) corresponding to
   the shuttling motion of a fluxon in the 1D case. The velocities 
   $v_I$ are indicated in the top right corner of the picture.}
\label{zs}
\end{figure}

Fig. \ref{zs} shows the regions of existence of ZFS in the $(L_I,C_I)$ plane. 
Notice the symmetry. For $C_I >> 1$ the solution is static 
while it is dynamic for $L_I>> 1$, this latter parameter 
plays a predominant role due to the interface conditions.

   \section{2D window junction }

In this section we discuss the dynamics of the kink in the window junction
and how to reduce it to a 1D effective problem. It is known that for a pure
Josephson junction this reduction is possible. For example in the static case
the behavior of a junction of width $w < 2 \pi$ is very well approximated
by the equation for the zero order mode \cite{cfgv96}.
For the dynamical case Eilbeck et al.
\cite{eilbeck} found that when $(I/{8 {\cal L}} \ll  1)$ the phase is uniform in 
the $y$ direction so that the reduction is possible. For a window junction
the situation is different because we have a passive region around the
junction which has electrical properties, a capacity and
an inductance which fix the velocity $v_{I}$. 
The comparison between 2D and 1D was done for the two cases of homogeneous
electrical properties in different geometries and a fixed geometry with
different electrical properties. Before giving these results we discuss
how the kink velocity is fixed by the presence of the passive region.
 
   \subsection {Justification of the effective 1D model}

When the velocity in the passive region $v_{I} \ne 1$  
the formula (\ref{v1d}) is completely off for the 2D calculations and the
reduction of the 2D window junction problem to a 1D effective problem 
is impossible.\\ 
Here the lateral passive region where waves can only propagate at
velocity $v_{I}$ is controlling the kink motion. The 
velocity -a free parameter for the sine-Gordon equation- can then
adjust itself to $v_{I}$ so that the "dressed" kink can propagate
in the device (junction and lateral passive region).
The fluxon adapts its velocity to $v_{I}$ so that $v=v_{I}$. 

To justify this, we take the time derivative of the first interface 
condition (\ref{ic}) and obtain
\begin{equation}
\label{tinter}
\phi_t = \psi_t ~~~{\rm on}~~~ \partial \Omega_J \;, 
\end{equation}
the second interface condition (\ref{ic}) reads 
\begin{equation}
\label{inter2}
\nabla \phi \cdot n = \nabla \psi \cdot n
\quad {\rm on} \quad \partial \Omega_J \;.
\end{equation}
The numerical simulations show that it is legitimate to 
assume that the solution is a 1D kink travelling in the passive region with 
the velocity $v_{I}$ and in the junction with velocity $v$.
Then  (\ref{inter2}) becomes at the longitudinal boundaries
\begin{equation}
\label{inter1d}
\partial_x \phi = \partial_x \psi ~~~{\rm on}~~~ \partial \Omega_J \;.
\end{equation}
The velocity $v$ of the kink on the boundary of the junction 
is defined by
\begin{equation}
\label{vj}
v=- \frac {\phi_t}{\phi_x} ~~~{\rm on}~~~ \partial \Omega_J \;.
\end{equation}
and the same for the velocity $v_{I}$ of the linear wave on the boundary 
of the junction 
\begin{equation}
\label{vidle}
v_{I}=- \frac {\psi_t}{\psi_x} ~~~{\rm on}~~~ \partial \Omega_J \;.
\end{equation}
Thus from (\ref{tinter}) and (\ref{inter1d}) we obtain $v=v_{I}$ so that
the estimate for the average voltage is then
\begin{equation} \label{v2d}
 V_{\rm 2D} = \frac {2\pi}{2w'/v_{I} + l/v_{I}} 
\end{equation} 

\begin{figure}[h]
\centerline{\psfig{figure=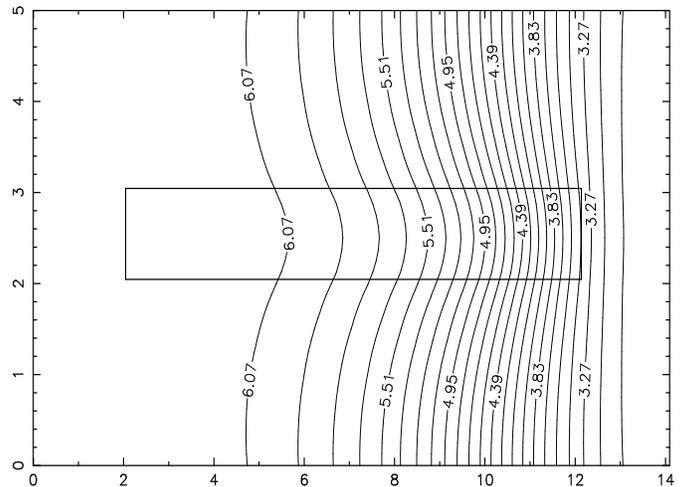,angle=-90,width=7cm,height=7cm}}
\caption{Contour of the phase for a homogeneous window junction with
$w'=2$ and $I=0.22$}
\label{contour1}
\end{figure}

   \subsection { Influence of the geometry on the zero field steps}

In this case the electrical properties are homogeneous in the whole device
so that ($L_I=C_I=1$). The numerical simulations show that for different
$w'$ the solution of the 2D problem are $y-$ uniform. For $w'=2$ for example
the Fig. \ref{contour1} represent the contour of the phase in the whole
domain at the time $T=5.024\,10^4$. We note that the Fig. \ref{contour1}
was normalized by ${\rm max}|\phi|$. The levels lines are parallel in the $y$
direction so that the solution is very close to a 1D kink. This 1D kink
propagates with a velocity $v_{I}$.

\begin{figure}[h]
\centerline{\psfig{figure=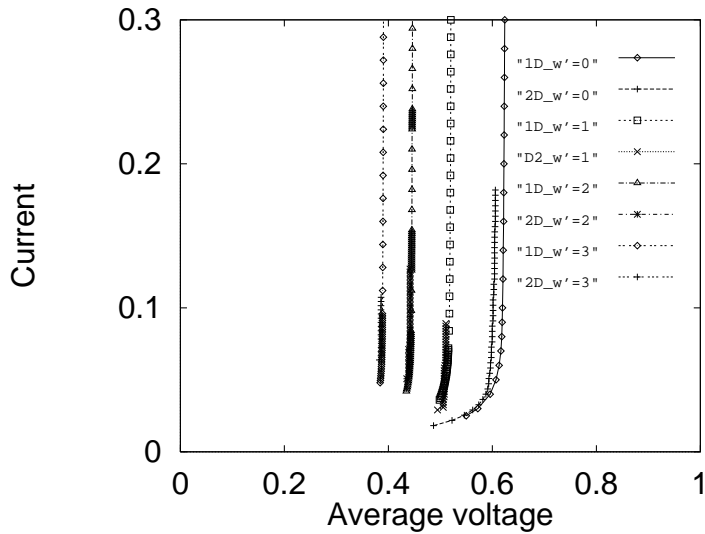,angle=-90,width=10cm,height=7cm}}
\caption{Current voltage characteristic for a homogeneous window junction
and various extensions of the passive region $w'$. Both 1D and 2D
calculations are presented for comparison.}
\label{iv5}
\end{figure}

The figure (\ref{iv5}) shows the IV characteristics for the both 1D and 2D
problems. One can see the good agreement between the 1D and 2D calculations.
Table (\ref{tab3}) reports the limiting values of the zero field step
computed numerically and the corresponding estimates from (\ref{v1d}) in excellent 
agreement.\\ 
As the numerical results show, in this case the dynamics of the solution
in the 2D window problem can be described by a unidimensional fluxon and 
the reduction to a 1D effective problem is possible.

\begin{table}[h]
\caption{positions of the ZFS for a both 1D and 2D homogeneous
junction  together with the estimates (\ref{v1d}).}
\vspace{0.2cm}
\begin{center}
\footnotesize
\begin{tabular}{||c|c|c|c||}
\hline  $w'$ & $V_{\rm 1D}$ & $V_{\rm num}$1D & $V_{\rm num}$2D\\
\hline  $0$ & $0.628$ & $0.625$  & $0.606$ \\
\hline  $1$ & $0.523$ & $0.522$ & $0.510$ \\
\hline  $2$ & $0.449$ & $0.448$ & $0.442$ \\
\hline  $3$ & $0.392$ & $0.391$ & $0.386$ \\ \hline
\end{tabular}
\end{center}
\label{tab3}
\end{table}

   \subsection {Influence of the electrical parameters }

We have obtained the same average voltage
for different values of $L_I\,C_I$ as long as the product $L_I~C_I= v_I^{-2}$
is kept constant.
We observe that the position of the zero
field step moves from the right to the left as the velocity in the
passive region $v_{I}$ decreases from $2$ to $0.5$.
We expect the ZFS to disappear when $v_{I}$ goes to $0$ ie $L_I$ or $C_I$
go to infinity.

   \subsection{ Instabilities}

As in 1D case there are two types of instabilities of the kink. 
The first is due to the geometry of the window, because large widths
$w'$ of the passive region (typically $w'=7$) lead
to a "stretched" kink in the junction which strongly radiates.
It is the impossible to accelerate such an object and have a stable
zero field step \cite{bc00}. This solution which is almost uniform
in the junction will oscillate but not rotate and the average voltage
will be zero \cite{bc00}. This situation was observed in the experimental work 
of Thyssen et al. \cite{Capri} who never found any zero field 
step for $w'/w < 3$.

\begin{figure}[h]
\centerline{\psfig{figure=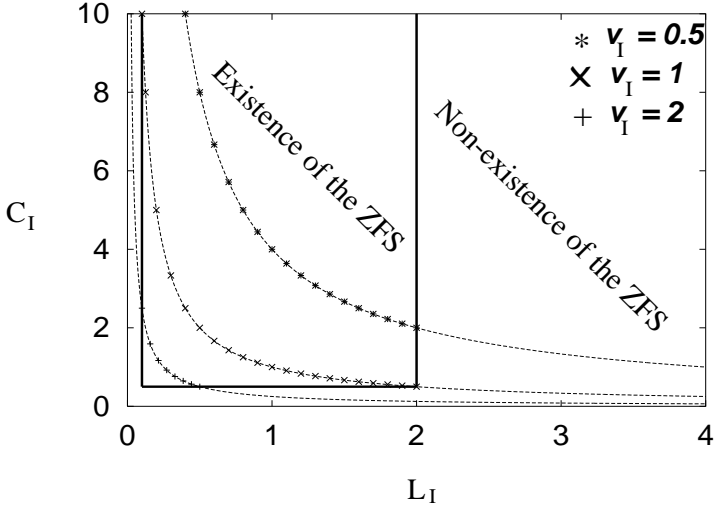,angle=-90,width=10cm,height=7cm}}
\caption{Parameter plane $(L_I, C_I)$ showing the regions
   of existence of zero field steps (ZFS) corresponding to
   the shuttling motion of a fluxon for 2D case. The velocities $v_I$ are indicated
   in the top right corner of the picture. The parameter  $w'=2$}
\label{zs2d}
\end{figure}

The second type of instability is due to the electrical quantities $L_I$ and
$C_I$. For $w'=2$, we have investigated the $(L_I, C_I)$ plane and found
numerically the regions of instability of the kink as shown in 
Fig. \ref{zs2d}.
We also plot three hyperbolas corresponding to $v_{I}= 0.5 (*), 1 (\times)$
and $2 (+)$. The markers indicate the positions where we have 
found the ZFS. All the points for a given marker correspond to the same
voltage. In Fig. \ref{zs2d} we have isolated the stability region as
the interior of the domain bounded by the solid line. We observe that the
instability is due essentially to the inductance $L_I$ so that the kink
motion is stable for $0.1 \le L_I\le 2$, independently of $C_I$. For
example $C_I = 10^4$ and $L_I=1$ gives rise to a ZFS. We can estimate
the average voltage using formula (\ref{v2d}), 
($V_{\rm 2D}= 4.5 \times 10^{-3}$ which is very close to the value
found numerically $V_{\rm num} 2D=4.577 \times 10^{-3}$). Thus
even for such a large value of $C_I$, the formula (\ref{v2d}) gives 
a good agreement with the numerical calculation.\\
Small values of $C_I$ lead to a static solution as can be seen from
equation (\ref{eqc4}) which reduces to $\Delta \psi=0$ so that 
$\psi$ is constant.
Because of the interface conditions $\phi$ is also constant.

\begin{figure}[h]
\centerline{\psfig{figure=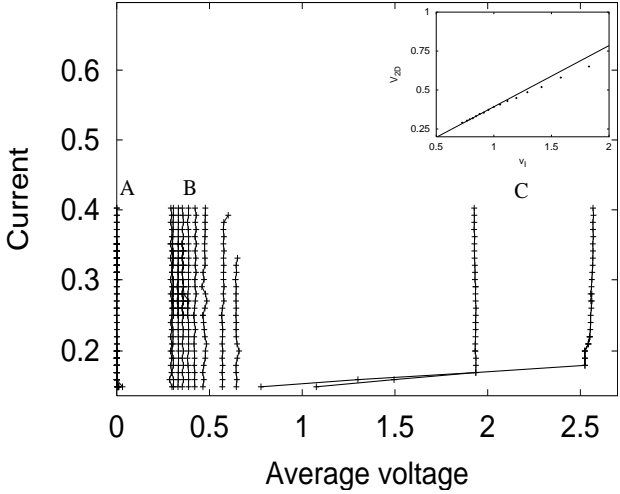,angle=-90,width=9cm,height=7cm}}
\caption{IV characteristics for $0.1 \le L_I\le 3$, $C_I=1$ and $w'=3$. The
insert shows the fluxon velocity in region B as a function of the linear
wave speed $v_I$ in the passive region.} 
\label{wp3li}
\end{figure}

\begin{figure}[h]
\centerline{\psfig{figure=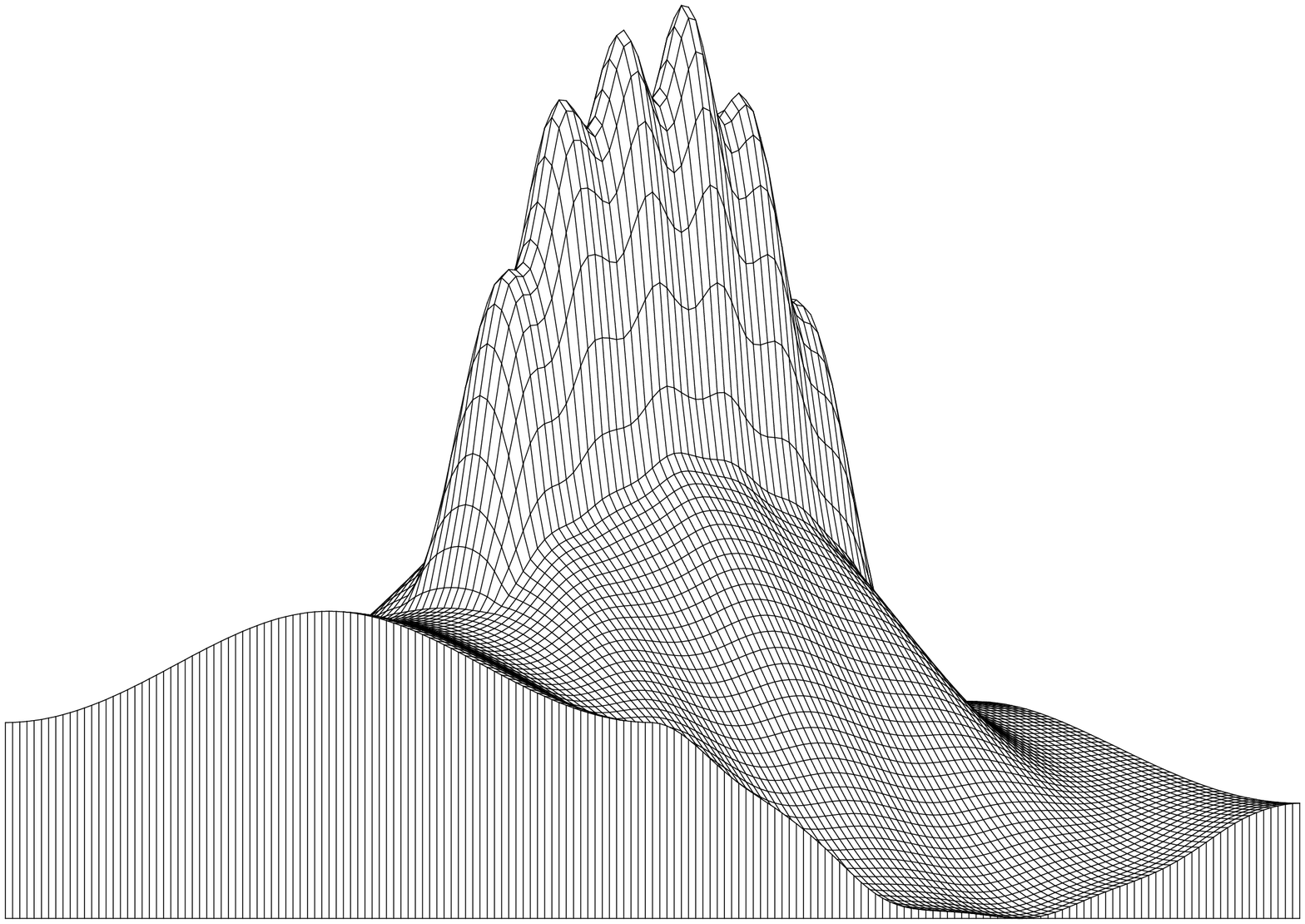,angle=-90,width=8.5cm,height=6cm}}
\centerline{ \psfig{figure=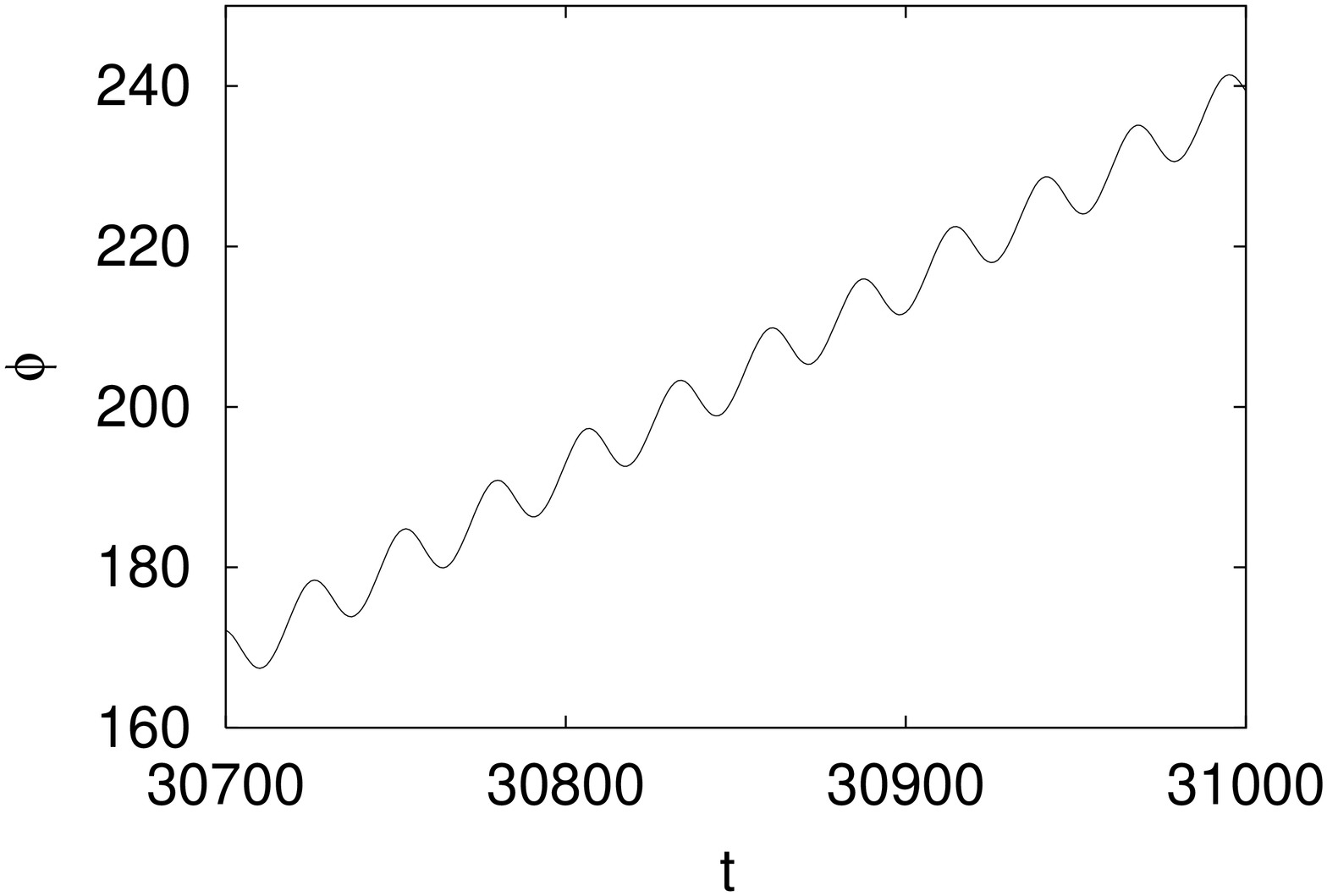,angle=-90,width=9cm,height=6cm}}
\caption{Three dimensional view of the phase at a given time (t=30800.) for 
$L_I=0.1$, region C in the previous picture (top panel). The bottom panel 
gives the temporal behavior of the phase in the middle of the junction.}
\label{li01}
\end{figure}

The influence of the inductance $L_I$ has been analyzed in detail
by calculating IV characteristics for $w'=3$ and $0.1 \le L_I\le 3$
which are plotted in Fig. \ref{wp3li}. There are three main regions
identified by the letters A B and C which correspond to three different
dynamical behaviors. In region A $2 \le L_I\le 3$, we obtain a static solution
and a zero voltage. At this time we do not understand why the kink
motion becomes unstable for $ L_I \ge 2$ independently of $C_I$.
This effect is present only in two dimensions and therefore linked to
the presence of a lateral passive region. It may be due to the large
jump in the gradient $\phi_y = (1/L_I) \psi_y$ along the lateral interface. \\
Region B corresponds to a ZFS thus a stable kink motion in the device. The
limiting voltage is given to a very good approximation by formula (\ref{v2d})
as shown by the inset of Fig. \ref{wp3li}. This expression also
gives a good estimate even for $v_I >1$. Region C $L_I=0.1,0.2$ corresponds
to much larger values of the voltage and a straight behavior typical of linear
resonances in the junction cavity. Small values of $L_I$ give rise to
a spatially uniform phase in the passive region, which reacts to the
phase in the junction. The voltage for $L_I=0.1$ corresponds to a higher
order cavity mode $V = 2.56 \approx 6 \pi / 10$. Fig. \ref{li01} 
shows the corresponding phase at a given instant (top panel) together
with the temporal behavior (bottom panel). As expected the phase is almost
uniform, the bounds in Fig. \ref{li01} are 191.83 and 193.14. The phase
increases linearly with time.

Then we conclude that the region of stability of the kink in the window
problem is
\begin{equation}
\label{reg_stab}
D_{\rm stability}=\lbrace 0.1 \leq L_I \leq 2 ~~~{\rm and}~~~ C_I \geq 0.5
\rbrace  \, .
\end{equation}

\section{Conclusion}

In this paper we have investigate the electrodynamics of a uniform
window junction, using the limiting case termed 1D model where the
lateral extension of the passive region is neglected.
We use the solution of the 1D model as a guide to discuss 
the behavior of the full window junction and this enables us 
to give an estimate of the average voltage for different extensions $w'$
of the passive region and different electrical parameters which fix
the velocity in the passive region $v_{I}$.
This estimate is in excellent agreement with the numerical results 
even when $v_{I} \neq 1$. In this situation we show that the 
fluxon adapts its velocity to $v_{I}$. 
For both cases $v_{I} =1 $ and $v_{I} \neq 1$ the numerical results show
that the fluxon propagates perpendicularly to the direction of the junction.
We have doubled the junction length to $l=20$ and find the same estimates
for the limiting voltage. In the 2D window junction, the 
translational symmetry is broken so that we obtain a different result
from the case where only a lateral passive region exists.

We have also studied numerically the stability of the zero field
step in the window junction and show that when the width of the 
passive part becomes large the fluxon becomes distorted and 
gives to another type of solution which is radial. 
In this case we have not found any ZFS.

This study shows that the kink can travel into the passive region
even when the impedance is not adapted as long as the mismatch
is in the capacitance and not the inductance. In other words the limit 
$\epsilon \rightarrow 0$ is not a singular limit. On the other hand
if there is an inductance mismatch then this will most likely cause
the break-up of kink shuttling leading to the disappearance of the
zero field step. The stability of the fluxon depends essentially on
the inductance $L_I$. This result, we believe, can lead to improving 
Josephson devices and their coupling to micro-strip lines.

 {\bf Acknowledgements}
   It is with great pleasure that J. G. C. thanks N. Flytzanis 
   for many useful discussions at the beginning of this work.
   A. B. thanks the Laboratory of Mathematics of the INSA de Rouen 
   and the Laboratoire de Physique th\'eorique de
   l'Universit\'e de Cergy-Pontoise for support. A. B. and J. G. C. thank
   Vladislav Kurin and Alexei Ustinov for many useful electronic 
   exchanges. We thank J\'er\^ome Dorignac for useful discussion.  
   This work was supported by the European Union under the RTN 
   project LOCNET HPRNCT-1999-00163. 
%
%

\vspace{.2in}

\appendix
\section{Derivation of the linear part of the $IV$ curve}

\vspace{.2in}

To determine the Hamiltonian associated with the 1D-window junction equation we multiply 
Eq. (\ref{eq1d}) by $ \partial \varphi/\partial t$ and integrate between $0$
and ${\cal L}$ to obtain
\begin{eqnarray}
\nonumber
{{\rm d} \over {{\rm d}t}}\left [\int_0^{\cal L}\frac{C(x)}2\varphi_t^2
+\epsilon(x)\left( 1-\cos \varphi \right)\;{\rm d}x \right] \\ \nonumber
-\int_0^{\cal L}\partial_x\left[{1 \over {L(x)}}\varphi_x\right]\varphi_t\;
{\rm d}x
= \int_0^{\cal L} \gamma (x) \varphi_t\;{\rm d}x - \\ \nonumber
\alpha\int_0^{\cal L} \epsilon(x) \varphi_t^2\;{\rm d}x \; ,
\nonumber
\end{eqnarray}
Integrating by parts the second term in the left-hand side and using the
boundary conditions we obtain
\begin{eqnarray}
\nonumber
{{\rm d} \over {{\rm d}t}}\left [\int_0^{\cal L}\frac{C(x)}2\varphi_t^2
+\frac 1{2L(x)}\varphi_x^2
+\epsilon(x)\left( 1-\cos \varphi \right)\;{\rm d}x \right]\\ \nonumber
= \int_0^{\cal L} \gamma(x) \varphi_t\;{\rm d}x 
 - \alpha\int_0^{\cal L}
\epsilon(x)\varphi_t^2\;{\rm d}x  \; .
\nonumber
\end{eqnarray}
Recalling the sine-Gordon-wave Hamiltonian 
\begin{equation}
\label{hsgw}
H=\int_0^{\cal L} \left [ \frac{C(x)}2\varphi_t^2+\frac
1{2L(x)}\varphi_x^2+\varepsilon (x)\left( 1-\cos \varphi \right) \right]
\;{\rm d}x
\end{equation}
The computation of the linear part of the $IV$ curve is based on the conservation 
of the energy of the system that is given by sG-wave Hamiltonian (\ref{hsgw})
from which it can be shown that $H$ satisfies the {\it power balance equation}
\begin{eqnarray}
{{\rm d} H \over {{\rm d}t}} = \gamma \int_0^{\cal L} \varphi_t\;{\rm d}x
-\alpha \int_{w'}^{w'+l} \varphi_t^2 \;{\rm d}x
\end{eqnarray}
In the stationary regime the average $dH/dt =0$ , so
$$
\left < \gamma \int_0^{\cal L} \varphi_t\;{\rm d}x \right > -
\alpha \left < \int_{w'}^{w'+l} \varphi_t^2 \;{\rm d}x \right > =0\, .
$$
\noindent
Since the voltage $V$ is the temporal average of $\varphi_t$, we obtain in
the high voltage limit $\left < \varphi_t^2 \right >  =  \left < \varphi_t \right >^2$, 
so that the IV curve is given by 
\[
V = \frac {\gamma {\cal L}} {\alpha l}
\]   
\bibliography{bc02}
\end{document}